\documentclass[aps,prd,twocolumn,superscriptaddress,nofootinbib]{revtex4-1}

\usepackage{graphicx,color}
\usepackage{latexsym,amsmath,amssymb,graphicx,booktabs}
\usepackage{hyperref}
\definecolor{MyBlue}{rgb}{0.15,0.15,0.70}
\hypersetup{
colorlinks=true,
citecolor=MyBlue,
linkcolor=MyBlue,
urlcolor=MyBlue
}

\newcommand{\iBox}{\Box^{-1}}

\newcommand{\Fmn}{F_{\mu\nu}}

\renewcommand\){\right)}

\renewcommand\]{\right]}

\def\lsim{\raise 0.4ex\hbox{$<$}\kern -0.8em\lower 0.62
ex\hbox{$\sim$}}

\def\gsim{\raise 0.4ex\hbox{$>$}\kern -0.7em\lower 0.62
ex\hbox{$\sim$}}

\def\lbar{{\hbox{$\lambda$}\kern -0.7em\raise 0.6ex
\hbox{$-$}}}

\newcommand\eq[1]{eq.~(\ref{#1})}

\newcommand\Eq[1]{Equation~(\ref{#1})}

\newcommand\pa{\partial}
\newcommand\p{\partial}

\newcommand\ee{\end{equation}}
\newcommand\be{\begin{equation}}
\def\bea{\begin{array}}
\def\eea{\end{array}}\def\ea{\end{array}}
\newcommand\ees{\end{eqnarray}}
\newcommand\bees{\begin{eqnarray}}

\def\d{\delta}

\def\dslash{\hspace{-1mm}\not{\hbox{\kern-2pt $\partial$}}}
\def\Dslash{\not{\hbox{\kern-2pt $D$}}}
\def\pslash{\not{\hbox{\kern-2.1pt $p$}}}
\def\kslash{\not{\hbox{\kern-2.3pt $k$}}}
\def\qslash{\not{\hbox{\kern-2.3pt $q$}}}

\newcommand{\vk}{{\bf k}}

\def\p1{{\bf p}_1}
\def\p2{{\bf p}_2}
\def\k1{{\bf k}_1}
\def\k2{{\bf k}_2}

\newcommand{\dddM}{\kern 0.2em \raise 1.9ex\hbox{$...$}\kern -1.0em \hbox{$M$}}
\newcommand{\dddQ}{\kern 0.2em \raise 1.9ex\hbox{$...$}\kern -1.0em \hbox{$Q$}}
\newcommand{\dddI}{\kern 0.2em \raise 1.9ex\hbox{$...$}\kern -1.0em\hbox{$I$}}
\newcommand{\dddJ}{\kern 0.2em \raise 1.9ex\hbox{$...$}\kern-1.0em
\hbox{$J$}}
\newcommand{\dddcalJ}{\kern 0.2em \raise 1.9ex\hbox{$...$}\kern-1.0em
\hbox{${\cal J}$}}

\newcommand{\dddO}{\kern 0.2em \raise 1.9ex\hbox{$...$}\kern -1.0em
\hbox{${\cal O}$}}
\def\dddz{\raise 1.5ex\hbox{$...$}\kern -0.8em \hbox{$z$}}
\def\dddd{\raise 1.8ex\hbox{$...$}\kern -0.8em \hbox{$d$}}
\def\dddbd{\raise 1.8ex\hbox{$...$}\kern -0.8em \hbox{${\bf d}$}}
\def\ddbd{\raise 1.8ex\hbox{$..$}\kern -0.8em \hbox{${\bf d}$}}
\def\dddx{\raise 1.6ex\hbox{$...$}\kern -0.8em \hbox{$x$}}

\newcommand{\Sch}{Schwarzschild }

\newcommand{\mplr}{m_{\rm Pl}}

\newcommand{\oma}{\Omega_{M}}
\newcommand{\ora}{\Omega_{R}}

\newcommand{\rde}{\rho_{\rm DE}}
\newcommand{\wde}{w_{\rm DE}}

\graphicspath{ {./Graphs/} }

\begin{document}


\title{The gravitational-wave luminosity distance\\ in modified gravity theories
}


\author{Enis Belgacem}
\affiliation{D\'epartement de Physique Th\'eorique and Center for Astroparticle Physics, Universit\'e de Gen\`eve, 24 quai Ansermet, CH--1211 Gen\`eve 4, Switzerland}

\author{Yves Dirian}
\affiliation{D\'epartement de Physique Th\'eorique and Center for Astroparticle Physics, Universit\'e de Gen\`eve, 24 quai Ansermet, CH--1211 Gen\`eve 4, Switzerland}

\author{Stefano Foffa}
\affiliation{D\'epartement de Physique Th\'eorique and Center for Astroparticle Physics, Universit\'e de Gen\`eve, 24 quai Ansermet, CH--1211 Gen\`eve 4, Switzerland}

\author{Michele Maggiore}
\affiliation{D\'epartement de Physique Th\'eorique and Center for Astroparticle Physics, Universit\'e de Gen\`eve, 24 quai Ansermet, CH--1211 Gen\`eve 4, Switzerland}



\begin{abstract}
In modified gravity   the propagation of gravitational waves (GWs)  is in general different from that in general relativity. As a result,
the luminosity distance  for  GWs  can differ from that for electromagnetic signals, and is affected both by the dark energy   equation of state $\wde(z)$ and by a function $\delta(z)$ describing modified propagation.
We show that the effect of modified propagation in general dominates over the effect of the dark energy   equation of state, making it easier to distinguish a modified gravity model from $\Lambda$CDM.
We illustrate this  using a nonlocal modification of gravity that has been shown to fit remarkably well CMB, SNe, BAO and structure formation data, and we discuss the prospects for distinguishing  nonlocal gravity from $\Lambda$CDM with  the Einstein Telescope. We find that, depending on the exact sensitivity, a few tens of standard sirens with measured redshift at $z\sim 0.4$, or
a few hundreds   at $1\,\lsim\,  z\, \lsim\, 2$, could suffice.

\end{abstract}


\maketitle

\section{Introduction}

The  observation  of the GWs from the neutron star binary coalescence 
GW170817~\cite{TheLIGOScientific:2017qsa} and of the associated $\gamma$-ray burst 
GRB~170817A  \cite{Goldstein:2017mmi,Savchenko:2017ffs,Monitor:2017mdv} has marked the opening of   the era of multi-messenger astronomy. In the near future more events of this type are expected, while, on a time-scale of 1-2 decades, the space interferometer LISA~\cite{Audley:2017drz} and a third-generation ground-based interferometer such as the Einstein Telescope (ET)~\cite{Sathyaprakash:2012jk} could extend these observations to large redshifts.

One of the most interesting targets of third-generation detectors  is the measurement of the luminosity distance  with  standard sirens~\cite{Schutz:1986gp,Dalal:2006qt,MacLeod:2007jd,Nissanke:2009kt,Cutler:2009qv,Sathyaprakash:2009xt,Zhao:2010sz,DelPozzo:2011yh,Nishizawa:2011eq,Taylor:2012db,Camera:2013xfa,Tamanini:2016zlh,Caprini:2016qxs,Cai:2016sby}. Currently, all  the studies on the subject have been performed using the standard expression of the luminosity distance in a theory with a dark energy (DE) density $\rde(z)$, 
\be\label{dLem}
d_L(z)=\frac{1+z}{H_0}\int_0^z\, 
\frac{d\tilde{z}}{E(\tilde{z})}\, ,
\ee
where
\be\label{E(z)}
E(z)=\sqrt{\ora (1+z)^4+\oma (1+z)^3+\rde(z)/\rho_0 }\, ,
\ee
and, as usual, $\rho_0=3H_0^2/(8\pi G)$ and $\ora$ and $\oma$ are the radiation and matter density fractions, respectively. The evolution of the DE density is determined by its equation of state (EoS) function $\wde(z)$ through the conservation equation
\be
\dot{\rho}_{\rm DE}+3H(1+\wde)\rde=0\, .
\ee 
Then, all works on cosmological applications of standard sirens either choose a simple phenomenological parametrization of $\wde(z)$, such as the $(w_0,w_a)$ parametrization 
$w_{\rm DE}(a)= w_0+(1-a) w_a$ \cite{Chevallier:2000qy,Linder:2002et} and provide forecasts on the accuracy to which $(w_0,w_a)$ can be measured, or develop methods for attempting a model-independent reconstruction of the function $\wde(z)$.

The most natural motivation for a  non-trivial dark energy EoS  is the assumption that gravity is modified at cosmological scales. 
Here we  point out, through the study of an explicit model, that in a generic modified gravity theory \eq{dLem} is not necessarily the correct luminosity distance for GWs (see also  
\cite{Deffayet:2007kf,Saltas:2014dha,Lombriser:2015sxa,Nishizawa:2017nef,Arai:2017hxj,Amendola:2017ovw}),
and we further show that the difference between the GW luminosity distance $d_L^{\,\rm gw}$ and the standard electromagnetic luminosity distance $d_L^{\,\rm em}$ gives an effect  that can be significantly larger than that due to a non-trivial dark energy EoS.

\section{Tensor perturbations in modified gravity} 

Let us first recall that, in GR,
  the free propagation of tensor perturbations  in a Friedmann-Robertson-Walker (FRW) background is described by 
\be\label{4eqtensorsect}
\tilde{h}''_A+2{\cal H}\tilde{h}'_A+k^2\tilde{h}_A=0\, ,
\ee
where $\tilde{h}_A(\eta, \vk)$ are  the Fourier modes of the GW amplitude, $A=+,\times$ labels the two polarizations, $\eta$ denotes  conformal time, the prime denotes $\pa_{\eta}$, and  
${\cal H}=a'/a$. Introducing  a field $\tilde{\chi}_A(\eta, \vk)$ from
\be\label{4defhchiproofs}
\tilde{h}_A(\eta, \vk)=\frac{1}{a(\eta)}  \tilde{\chi}_A(\eta, \vk)\, ,
\ee
\eq{4eqtensorsect} becomes
\be
\tilde{\chi}''_A+\(k^2-a''/a\) \tilde{\chi}_A=0\, .
\ee
Both in matter dominance and in the recent DE dominated epoch  $a''/a\sim 1/\eta^2$. For sub-horizon modes $k\eta\gg 1$, and therefore 
$a''/a$ can be neglected compared to $k^2$. For   GWs observed at ground- or space-based interferometers this holds to huge accuracy: for instance, for a GW frequency  $f\sim 10^2$~Hz,
\be
(k\eta)^{-2}\sim (500\, {\rm km}/H_0^{-1})^2\sim 10^{-41}\, .
\ee 
Then, we can write simply
\be
\tilde{\chi}''_A+k^2 \tilde{\chi}_A=0\, .
\ee
This shows that the dispersion relation of tensor perturbations is $\omega=k$, i.e. GWs propagate at the speed of light (that we have set to one). On the other hand, the factor $1/a$ in \eq{4defhchiproofs}
tells us how the GW amplitude decreases in the propagation over cosmological distances from the source to the observer and, for inspiraling binaries, leads to the standard dependence of the GW amplitude $\tilde{h}_A(\eta, \vk)\propto 1/d_L(z)$; see e.g. 
Section 4.1.4 of \cite{Maggiore:1900zz}.

In a generic modified gravity theory both the coefficient of the $k^2$ term and that of the $2{\cal H}$ term in \eq{4eqtensorsect} (as well as the source term, that we have not written explicitly) can  be different. This has already been observed in various explicit models. In particular, in the  DGP model~\cite{Dvali:2000hr} (which, in the self-accelerated branch,  is by now ruled out by the presence of 
instabilities at the level of cosmological perturbations~\cite{Luty:2003vm,Nicolis:2004qq,Gorbunov:2005zk,Charmousis:2006pn}), at cosmological scales gravity  leaks into extra dimensions, and this affects the $1/d_L(z)$ behavior of a gravitational signal~\cite{Deffayet:2007kf}.  The same effect has been found  
 for  Einstein-Aether models and for scalar-tensor   theories of the Horndeski class~\cite{Saltas:2014dha,Lombriser:2015sxa,Arai:2017hxj,Amendola:2017ovw}.  A modified propagation equation for tensor modes can be included  in the general effective field theory approach to dark energy developed in \cite{Gleyzes:2014rba}, and the relevance of this effect for standard sirens has already been pointed out, in a scalar-tensor theory of the Horndeski class, in \cite{Lombriser:2015sxa}.\footnote{A general formalism for testing gravity with GW propagation has been recently presented in
\cite{Nishizawa:2017nef}. Ref.~\cite{Yunes:2016jcc} gives  a detailed discussion of the constraints obtained from the  first two observations of BH-BH coalescences, 
GW150914 and GW151226, both on  modified GW generation and   on  modified GW propagation due to a non-trivial  dispersion  relation of the graviton.}

A change in the coefficient of the $k^2$ term in \eq{4eqtensorsect} gives a propagation speed of GWs different from the speed of light. The GW170817/GRB~170817A  event now puts a very stringent limit on such a modification, at the level  $|c_{\rm gw}-c|/c< O(10^{-15})$ \cite{Monitor:2017mdv}, which rules out a large class of scalar-tensor and vector-tensor modifications of GR~\cite{Creminelli:2017sry,Sakstein:2017xjx,Ezquiaga:2017ekz,Baker:2017hug}. Let us then focus on the effect of modifying the coefficient of the 
$2{\cal H}$ term, i.e. let us consider a propagation equation of the form
\be\label{prophmodgrav}
\tilde{h}''_A  +2 {\cal H}[1-\delta(\eta)] \tilde{h}'_A+k^2\tilde{h}_A=0\, ,
\ee
with $\delta(\eta)$ some function (we will present in Section~\ref{sect:modpropNL} an explicit example of a modified gravity model where GW propagation is described by such an equation).
In this case we introduce $\tilde{\chi}_A(\eta, \vk)$ from 
\be\label{4defhchiproofsRR}
\tilde{h}_A(\eta, \vk)=\frac{1}{\tilde{a}(\eta)}  \tilde{\chi}_A(\eta, \vk)\, ,
\ee
where 
\be\label{deftildea}
\frac{\tilde{a}'}{\tilde{a}}={\cal H}[1-\delta(\eta)]\, ,
\ee
and we get $\tilde{\chi}''_A+(k^2-\tilde{a}''/\tilde{a}) \tilde{\chi}_A=0$. Once again, inside the horizon the term $\tilde{a}''/\tilde{a}$ is totally negligible, so   GWs propagate at the speed of light.
However, in the propagation across cosmological distances, $\tilde{h}_A$ now decreases as $1/\tilde{a}$ rather than $1/a$.
Then, in such a modified gravity model we must distinguish between an 
electromagnetic luminosity distance $d_L^{\,\rm em}(z)$ and a GW luminosity distance $d_L^{\,\rm gw}(z)$, and 
the GW amplitude  of a coalescing binary at redshift $z$ will now be proportional 
to $1/d_L^{\,\rm gw}(z)$, where
\be
d_L^{\,\rm gw}(z)=\frac{a(z)}{\tilde{a}(z)}\, d_L^{\,\rm em}(z)
=\frac{1}{(1+z)\tilde{a}(z)}\, d_L^{\,\rm em}(z)\, ,
\ee
and $d_L^{\,\rm em}(z)\equiv d_L(z)$ is the standard luminosity distance (\ref{dLem}) for electromagnetic signals. \Eq{deftildea} is equivalent to $(\log a/\tilde{a})'=\delta(\eta) {\cal H}(\eta)$, which is easily integrated  and gives 
\be\label{dLgwdLem}
d_L^{\,\rm gw}(z)=d_L^{\,\rm em}(z)\exp\left\{-\int_0^z \,\frac{dz'}{1+z'}\,\delta(z')\right\}\, .
\ee

\section{Modified propagation in nonlocal gravity} \label{sect:modpropNL}

To illustrate this effect, and the relative roles of $\wde(z)$ and $\d(z)$ in  $d_L^{\,\rm gw}(z)$, we  consider an explicit modified gravity model, but, as will be clear,  the results that we  find are more general. The model that we consider  is a nonlocal modification of gravity that has been introduced and much studied in recent years by our group. The underlying physical idea is that, even if the fundamental action of gravity is local, the corresponding quantum effective action, 
that includes the effect of quantum fluctuations, is  nonlocal. 
These nonlocalities are well understood in the ultraviolet regime, where their computation is by now standard textbook material~\cite{Birrell:1982ix,Mukhanov:2007zz,Shapiro:2008sf}, but are much less understood in the infrared (IR), which is the regime relevant for cosmology.  IR effects in quantum field theory in curved space have been studied particularly in de~Sitter space where strong effects, due in particular to the propagator of the conformal mode \cite{Antoniadis:1986sb}, have been found. However, the whole issue of IR corrections in de~Sitter space is unsettled, because of the intrinsic difficulty of the problem. Given the difficulty of a pure top-down approach, we have taken an alternative and more phenomenological strategy. In general, strong IR effects  manifest themselves through the generation of nonlocal terms, proportional to inverse powers of the d'Alembertian operator, in the quantum effective action. For instance, in QCD  the strong IR fluctuations generate a term~\cite{Boucaud:2001st,Capri:2005dy,Dudal:2008sp} 
\be\label{Fmn2}
\frac{m_g^2}{2} {\rm Tr}\, \int d^4x\,   F_{\mu\nu} \frac{1}{\Box}F^{\mu\nu}\, ,
\ee
in the quantum effective action, where $\Fmn$ is the non-abelian field strength. This nonlocal term corresponds to giving a mass $m_g$ to the gluons: indeed,   choosing the Lorentz gauge and expanding in powers of the gauge field $A^{\mu}$, the above terms gives  a gluon mass term $m_g^2 {\rm Tr} (A_{\mu}A^{\mu})$, plus extra  nonlocal interactions. Note that the use of a nonlocal operator such as $\iBox$ allows us to write a mass term without violating gauge invariance. However, this only makes sense at the level of quantum effective actions, where nonlocalities are unavoidably generated by quantum loops whenever the theory contains massless or light particles. The fundamental action of a quantum field theory, in contrast, must be local. Thus, nonlocal terms of this form describe dynamical mass generation by quantum fluctuations at the level of the quantum effective action.

In a similar spirit, we have studied a model of gravity based on the quantum effective action
\be\label{RR}
\Gamma_{\rm RR}=\frac{\mplr^2}{2}\int d^{4}x \sqrt{-g}\, 
\[R-\frac{1}{6} m^2R\frac{1}{\Box^2} R\]\, ,
\ee
where $\mplr$ is the reduced Planck mass and  $m$ is a  new mass parameter that replaces the cosmological constant of $\Lambda$CDM. This model was  proposed in \cite{Maggiore:2014sia}, following earlier work in \cite{Maggiore:2013mea}, and   it  can be shown that the nonlocal term in \eq{RR} corresponds to a dynamical mass generation for the conformal mode of the metric~\cite{Maggiore:2015rma,Maggiore:2016fbn}. Recently, some evidence for the nonlocal term in \eq{RR} has also been found from non-perturbative studies in lattice gravity~\cite{Knorr:2018kog}.
A  detailed comparison with cosmological data  and Bayesian parameter estimation has been carried out  in 
\cite{Dirian:2014ara,Dirian:2014bma,Dirian:2016puz,Dirian:2017pwp,Belgacem:2017cqo}, where it has been found that the model fits 
cosmic microwave background (CMB), supernovae (SNe), 
baryon acoustic oscillation (BAO), structure formation and local $H_0$ measurements  at a level statistically indistinguishable from $\Lambda$CDM (with the same number of parameters, since $m$ replaces $\Lambda$); furthermore, parameter estimation gives a large value of the Hubble parameter, which basically eliminates the tension  between the {\em Planck} CMB data~\cite{Planck_2015_CP} and
the  local $H_0$ measurements~\cite{Riess:2016jrr}. The parameter $m$ is also fixed by Bayesian parameter estimation from CMB, SNe and BAO data, and turns out to be of order $H_0$.
The model has been reviewed in~\cite{Maggiore:2016gpx} and, more recently, in \cite{Belgacem:2017cqo}, to which we refer the reader for  a  detailed discussion of  conceptual aspects and  phenomenological consequences. We will refer to it as the ``RR" model.

\begin{figure}[t]
\includegraphics[width=0.35\textwidth]{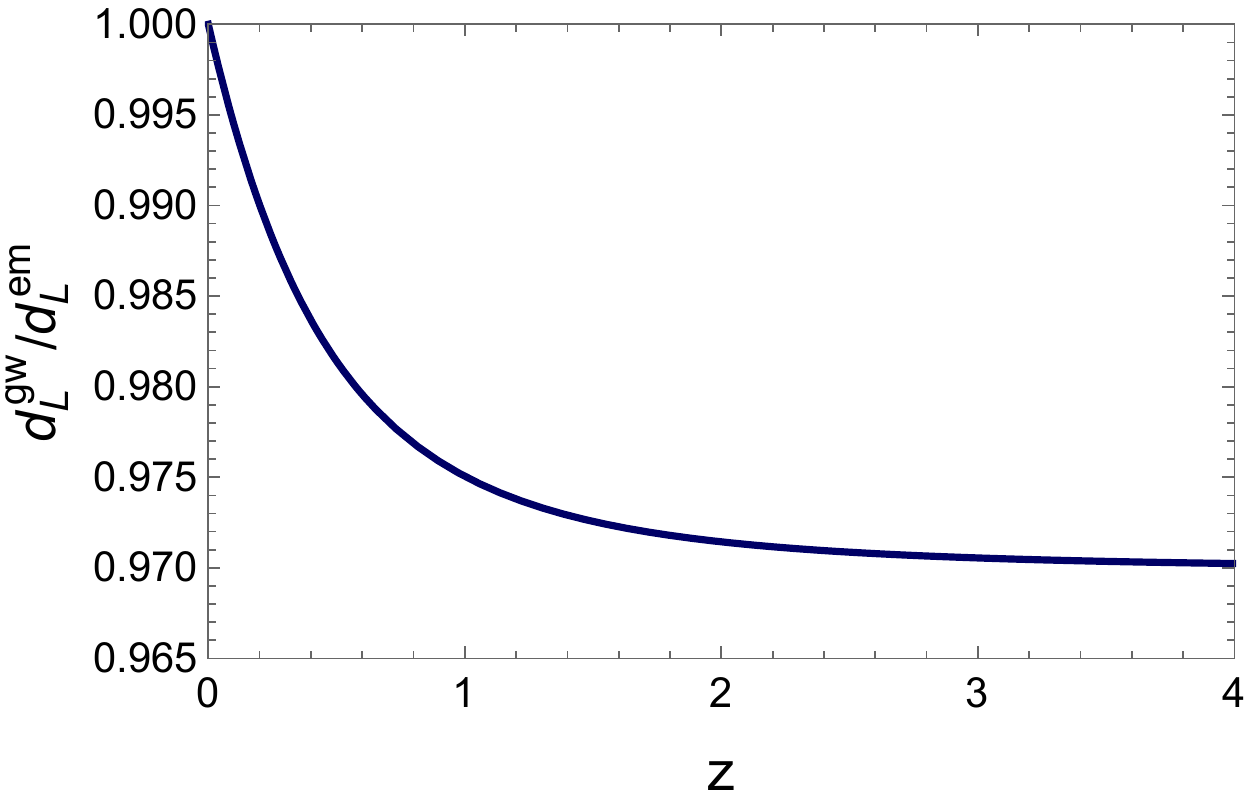}
\caption{The ratio $d_L^{\,\rm gw}(z)/d_L^{\,\rm em}(z)$ in the RR model.}
\label{fig:dLgw_over_dLem}
\end{figure}

\begin{figure}[t]
\includegraphics[width=0.35\textwidth]{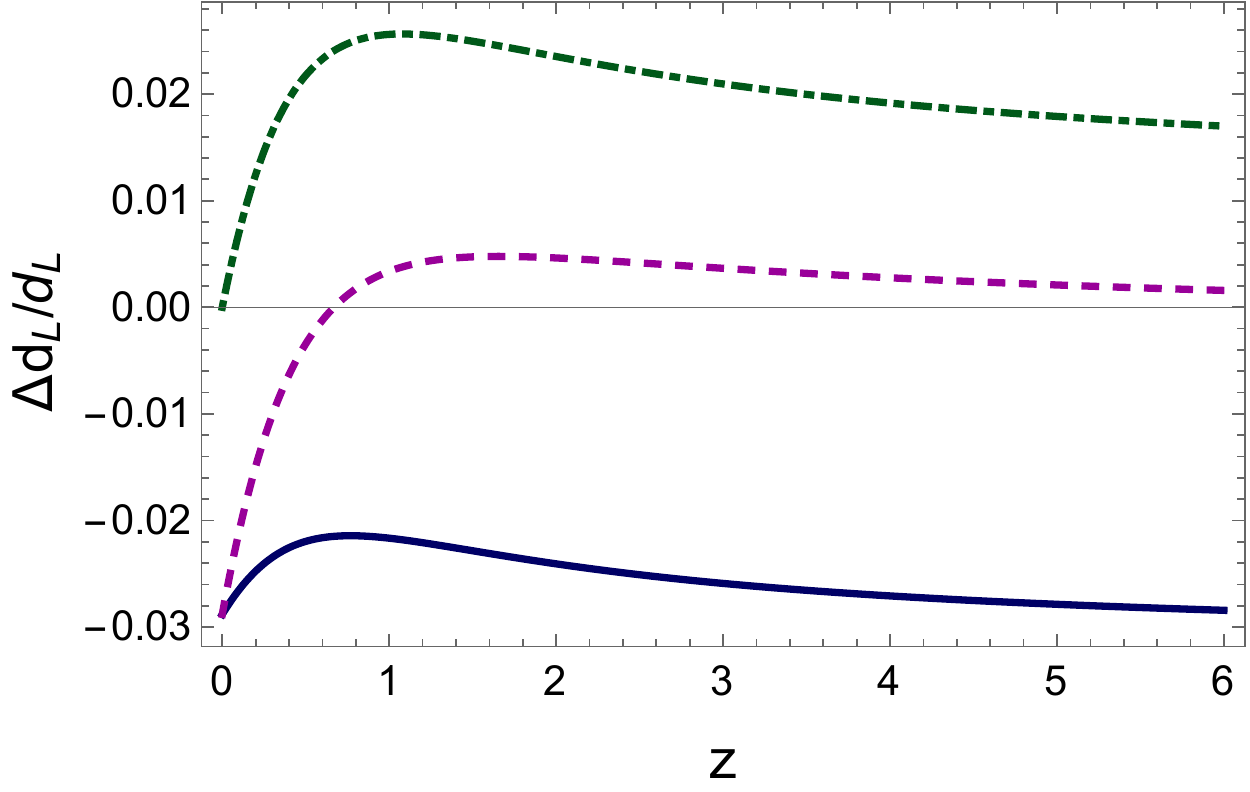}
\caption{The relative differences $\Delta d_L/d_L$ between the RR model and $\Lambda$CDM for three different cases. Green, dot-dashed curve: the relative difference 
$(d_L^{\rm RR,em}-d_L^{\Lambda{\rm CDM}})/d_L^{\Lambda{\rm CDM}}$ using the same values of $h_0$ and $\oma$ (taken for definiteness as $h_0=0.7013$ and $\oma=0.2922$).  Dashed magenta curve: the same, but using for each model its own mean values of  $h_0$ and $\oma$.  Blue solid line: the relative difference $\Delta d^{\rm gw}_L/d_L\equiv
(d_L^{\rm RR,gw}-d_L^{\Lambda{\rm CDM}})/d_L^{\Lambda{\rm CDM}}$
using again for each model its own mean values of $h_0$ and $\oma$.}
\label{fig:Deltad_over_d}
\end{figure}

The equation of tensor perturbations in the RR model has been derived in \cite{Dirian:2016puz} and, for the free propagation, has indeed the form (\ref{prophmodgrav}), with 
\be\label{defdeltaperRR}
\delta=\frac{3\gamma (d\bar{V}/d\log a)}{2(1 - 3\gamma \bar{V})}\, ,
\ee
where $\bar{V}$ is the background evolution of an auxiliary field that is introduced to rewrite \eq{RR} in local form (see e.g. Section~3 of \cite{Belgacem:2017cqo} for review), and $\gamma= m^2/(9H_0^2)$.    For this form of $\delta(z)$, the integral in  \eq{dLgwdLem} can be  computed analytically by transforming the integration over  $dz$ into an integration over $d\bar{V}$, which gives
\be\label{dLgwdLemRR}
d_L^{\rm RR,gw}(z)=d_L^{\rm RR,em}(z)\, \sqrt{\frac{1-3\gamma\bar{V}(0) }{1-3\gamma\bar{V}(z) }}\, ,
\ee
so in  the RR model the ratio $d_L^{\,\rm gw}(z)/d_L^{\,\rm em}(z)$ is a local function of $\bar{V}(z)$. We plot this ratio in
Fig.~\ref{fig:dLgw_over_dLem}. In the RR model,  scalar perturbations obey a modified Poisson equation with an effective Newton constant that, for modes well inside the horizon, is given by~\cite{Dirian:2014ara}
\be\label{Geffdiz}
G_{\rm eff}(z)=\frac{G}{1-3\gamma\bar{V}(z)}\, .
\ee 
Then, \eq{dLgwdLemRR} can be rewritten as
\be\label{dLgwdLemGeff}
d_L^{\rm RR,gw}(z)=d_L^{\rm RR,em}(z)\, \sqrt{\frac{G_{\rm eff}(z) }{G_{\rm eff}(0)}}\, ,
\ee
that nicely ties modified GW propagation  to the modification in the growth of structures.
Quite remarkably, this is exactly the same relation  found recently in a subclass of Horndeski models
\cite{Linder:2018jil}.
 
In Fig.~\ref{fig:Deltad_over_d} we  show the relative difference $\Delta d_L/d_L$ for three different cases.
The upper curve  is the relative difference 
 between the electromagnetic luminosity distance   in  the RR model and the luminosity distance of $\Lambda$CDM,  when we use the same fiducial values for  
$h_0$ 
and $\oma$. In this case we see that, over a range of redshifts relevant for third-generation interferometers, the relative difference is of order $2\%$. 
However, this is not the  quantity relevant to observations. For each model, the actual  predictions are those obtained  by using its own best-fit values (or the mean values, or the priors) of the cosmological parameters, which are found by comparing the model with a set of cosmological data and performing Bayesian parameter estimation.  For the RR model, as well as for $\Lambda$CDM,
this is obtained 
by computing the  cosmological perturbations of the  model, inserting them in a Boltzmann code, and constraining the model  with observations by using a Markov Chain Monte Carlo. For the RR model this has been done in \cite{Dirian:2014bma,Dirian:2016puz,Dirian:2017pwp,Belgacem:2017cqo}.
Here we will use for definiteness the values in Table~3 of \cite{Belgacem:2017cqo}, where we used as datasets the {\em Planck\,} CMB data, a compilation of BAO data, the SNe data from the JLA dataset, and the local measurement of $H_0$. In this case for $\Lambda$CDM  we get the mean values
$h_0=0.681(5)$  and $\oma =0.305(7)$, while for the ``minimal" RR model (in which a parameter $u_0$ that determines the initial condition of an auxiliary field is set to zero; the limit of large $u_0$ brings the model closer and closer to $\Lambda$CDM) we get $h_0=0.701(7)$ and
$\oma =0.292(8)$. 
The corresponding result for  $(d_L^{\rm RR,em}-d_L^{\Lambda{\rm CDM}})/d_L^{\Lambda{\rm CDM}}$ is given by the dashed, magenta curve in Fig.~\ref{fig:Deltad_over_d} and we see that, at redshifts $z\, \gsim\, 1$, is  one order of magnitude smaller than the green curve. This is easily understood. Parameter estimation is basically performed by comparing the predictions of each model to a set of fixed distance indicators, such as  those given by the peaks of the CMB or by the BAO scale. Thus, the parameters in each model are adjusted so to reproduce these distance measurements at large redshift, and therefore have the tendency to compensate the differences in luminosity distance (or in comoving distance or in angular diameter distance) induced by the different functional forms of $\wde(z)$. As a result, at redshifts $z\, \gsim\, 0.5$, $|\Delta d_L|/d_L$ is reduced by about one order of magnitude, to a value  $(0.2-0.4)\%$, which is  much more difficult to observe. 
It is clear, from the above physical explanation, that this effect is quite general in modified gravity models, and we have detected it in the RR model simply because in this case a detailed Bayesian parameter estimation was already available.
 
The two upper curves in Fig.~\ref{fig:Deltad_over_d} give
the relative difference of the {\em electromagnetic} luminosity distances, which is relevant for standard candles. For standard sirens we rather need to compare the GW luminosity distance of the RR model, $d_L^{\rm RR,gw}$, to the luminosity distance 
$d_L^{\Lambda{\rm CDM}}$ of $\Lambda$CDM (which, in contrast, is the same for GWs and for electromagnetic signals). The result of this comparison, using  again the respective mean values of the parameters for the RR model and for $\Lambda$CDM, is given by the lower curve (blue, solid line) in
Fig.~\ref{fig:Deltad_over_d}. We see that the difference, in absolute value, now raises again to values of order $3\%$, and  the sign of the difference is opposite.

From these results we can draw some interesting conclusions. First, the existence in generic modified gravity theories of a notion of  GW luminosity distance, a priori different from the electromagnetic luminosity distance, makes in principle possible a conceptually clean test of modifications of GR. If the luminosity distance derived from a set of standard candles   turns out to be different from the result obtained with standard sirens, this will be a ``smoking gun" evidence for  modified gravity (see also \cite{Saltas:2014dha,Lombriser:2015sxa,Nishizawa:2017nef}). A second  point is that, at the redshifts $z\, \gsim \, 1$ relevant for LISA and ET,  the deviation from
the $\Lambda$CDM prediction induced by $\d(z)$ is much larger than that induced by $\wde(z)$, i.e., in absolute value, in Fig.~\ref{fig:Deltad_over_d} the blue solid curve is larger than the magenta 
dashed curve.\footnote{Of course, in a given specific modified gravity model, the function $\delta (z)$ could simply be zero, or anyhow such that $|\delta(z)|\ll |1+\wde(z)|$, in which case the main effect would come from $\wde(z)$. What our argument shows is that, in a generic modified gravity  model where the deviation of $\d(z)$ from zero and the deviation of $\wde(z)$ from $-1$ are of the same order, the effect of $\d(z)$ dominates.}
Note also that, if one measures a deviation from $\Lambda$CDM of the type of the blue solid curve in Fig.~\ref{fig:Deltad_over_d} and tries to interpret it as due to a non-trivial dark-energy equation of state, neglecting the possibility of modified GW propagation,  one would conclude that this  is a signature of a non-phantom $\wde(z)$ (which results in a negative value for  $\Delta d_L/d_L$). However, this interpretation could be totally wrong. In our case, the blue solid curve in Fig.~\ref{fig:Deltad_over_d} is produced in a model, such as the RR model, that has a phantom DE equation of state, and  the effect is not due to $\wde(z)$, but is rather dominated by $\d(z)$.

\section{Comparison with the Einstein Telescope} 

In a generic modified gravity model there will be both differences in the propagation of GWs and in their production mechanism, compared to GR. The two effects are decoupled, the former affecting the luminosity distance, as we have seen, and the latter the phase of the GW signal. The modification to the production mechanism depends on how much the modified gravity theory differs from GR at the distance scale $L$ of the binary system (and on whether it contains extra radiative degrees of freedom). In the  RR model there are no extra radiative degrees of freedom, and the static  \Sch solution of the theory reduces smoothly to that of GR at distances $L\ll m^{-1}\simeq H_0^{-1}$, with corrections  of order $(mL)^2$ \cite{Kehagias:2014sda,Maggiore:2014sia}. For $L$ of the order of the size of an astrophysical binary this  correction is utterly negligible and does not affect GW production. A more subtle point is whether the time dependence (\ref{Geffdiz}) of the effective Newton constant, found on cosmological scales, can be extrapolated down to the scale of a coalescing binary (see \cite{Barreira:2014kra} and  Appendix~B of \cite{Dirian:2016puz} for  discussion). In any case, the effect on the waveform due to modifications of GW propagation in general dominates over modification of GW production, since the former gives an effect that accumulates over the distance to the source~\cite{Yunes:2016jcc}, and here we focus on it.

\begin{figure}[t]
\includegraphics[width=0.35\textwidth]{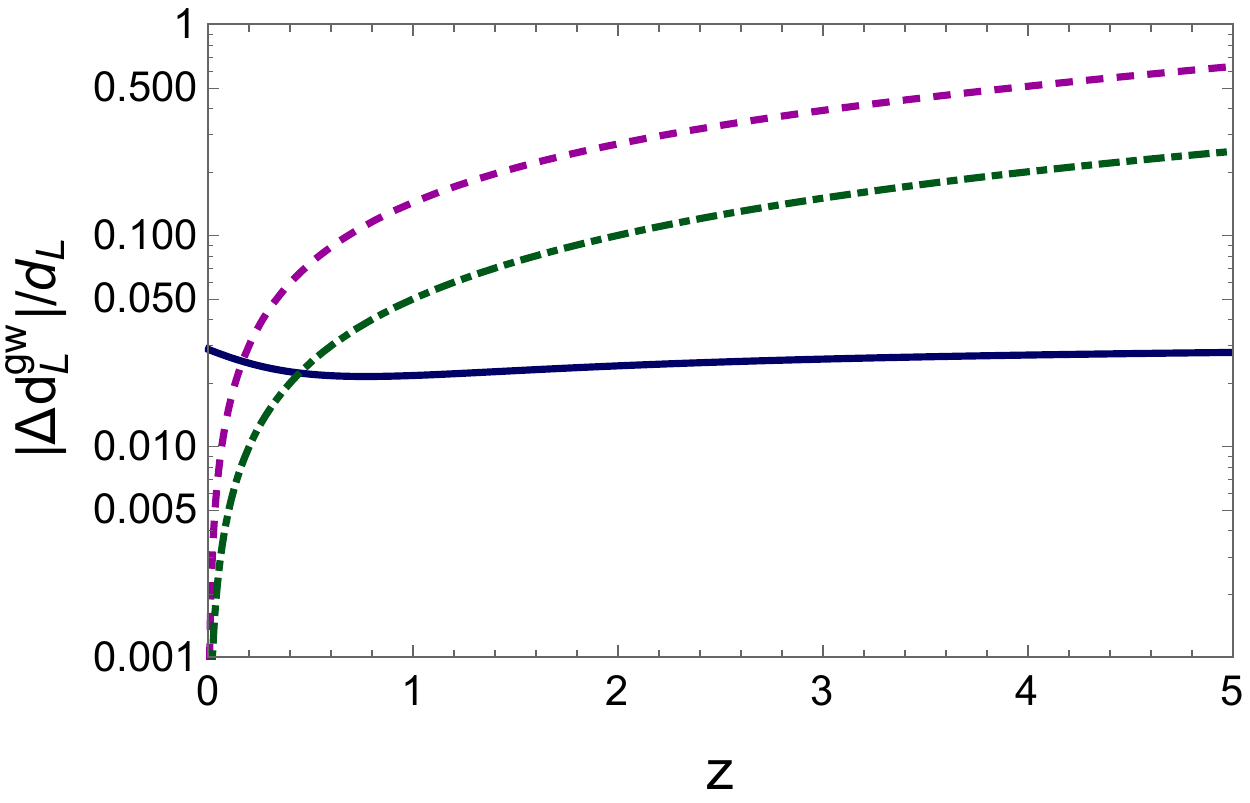}
\caption{The absolute value of $\Delta d^{\rm gw}_L/d_L\equiv (d_L^{\rm RR,gw}-d_L^{\Lambda{\rm CDM}})/d_L^{\Lambda{\rm CDM}}$, where both $d_L^{\rm RR,gw}$  and $d_L^{\Lambda{\rm CDM}}$ are computed using the respective mean values of the parameters (blue solid line), compared with an estimate of the total error on $\Delta d^{\rm gw}_L/d_L$ for ET (magenta, dashed) and the contribution to the error due to lensing (green, dot-dashed).}
\label{fig:comparisonErrorET}
\end{figure}

 In Fig.~\ref{fig:comparisonErrorET} we show  
\be
\left| \frac{\Delta d^{\rm gw}_L}{d_L}\right|\equiv
\frac{|d_L^{\rm RR,gw}-d_L^{\Lambda{\rm CDM}}|}{d_L^{\Lambda{\rm CDM}} }\, ,
\ee
and we compare it with an estimate of the total error  in ET due to instrumental noise 
plus lensing ~\cite{Zhao:2010sz}, and with the separate contribution to the error due to lensing ~\cite{Sathyaprakash:2009xt}. While the instrumental error  is inversely proportional to the signal-to-noise ratio of the source, and can in principle be decreased by improving the detector sensitivity, the error due to lensing is due  to  intervening  matter  structures that affect the GW propagation, and
provides a lower limit on the error of a third-generation interferometer (unless suitable delensing techniques are applied). 
Note that at very low redshifts, $z\, \lsim\, 0.05$, the error in the uncertainty in the local Hubble flow (not shown in the figure) will eventually dominate.  Given 
that with $N$ measurements we improve the accuracy by a factor $\sqrt{N}$, from this plot we find that,
to reach a sensitivity of the order of the signal, we need about 
7 standard sirens (with measured redshift) at $z\simeq 0.4$, or
about $44$ standard sirens at $z\simeq 1$, or $130$ at    $z\simeq 2$. Thus, a significant signal-to-noise ratio could be obtained with a few tens of standard sirens at $z\sim 0.5$, or a few hundreds at $z\sim 1-2$.
Of course, these numbers should only be taken  as indicative, since the sensitivity of third-generation interferometers is still quite tentative. 
Note also that current  errors on the estimate of cosmological  parameters such as $h_0$ and $\oma$ are of order $2\%$. This induces a corresponding theoretical error in the prediction, that  is not negligible compared to the predicted value of $|\Delta d^{\rm gw}_L/d_L|$. However, by the time  that third-generation interferometers will operate, further improvement in cosmological parameter estimation   is expected from mid-future observations
such as the EUCLID mission~\cite{Laureijs:2011gra}, DESI  \cite{DESI-1} or  SKA
\cite{Bull:2015nra}; otherwise, a larger number of sources will be necessary. In any case, 
a third-generation interferometer such as ET is expected to detect millions of binary mergers, of which possibly $O(10^3-10^4)$ could have an electromagnetic counterpart. Prospects for dark energy studies using standard sirens therefore look bright.

\vspace{5mm}\noindent
{\bf Acknowledgments.} 
The work  of the authors is supported by the Fonds National Suisse and  by the SwissMap National Center for Competence in Research.

\bibliography{myrefs_massive}

\end{document}